 \documentclass[twocol]{ametsocV6.1}
\usepackage{multirow}
\usepackage{physics}
\usepackage{array}

\bibpunct{(}{)}{;}{a}{}{,}

\title{A Model for the Tropical Cyclone Wind Field Response to Idealized Landfall\newline
\small\color{red} Submitted to Journal of the Atmospheric Sciences for peer review}

\authors{Jie Chen,\aff{a,b}\correspondingauthor{Jie Chen, chenjie@princeton.edu} 
Daniel R. Chavas,\aff{b} 
}

\affiliation{\aff{a}{Program in Atmospheric and Oceanic Sciences, Princeton University, Princeton, NJ}\\
\aff{b}{Dept. of Earth, Atmospheric and Planetary Sciences, Purdue University, West Lafayette, IN}\\
}

\abstract{
The impacts of a tropical cyclone after landfall depend not only on storm intensity but also on the size and structure of the wind field. Hence, a simple predictive model for the wind field after landfall has significant potential value. This work tests existing theory for wind structure and size over the ocean against idealized axisymmetric landfall experiments in which the surface beneath a mature storm is instantaneously dried and roughened individually or simultaneously. Structure theory captures the response of the low-level wind field to different types of idealized landfalls given intensity and size response. Storm size, modeled to follow the ratio of simulated time-dependent storm intensity to the Coriolis parameter $\frac{v_m(\tau)}{f}$, can generally predict the transient response of the storm gale wind radii $r_{34kt}$ to inland surface forcings, particularly for at least moderate surface roughening regardless of the level of drying. Given knowledge of the intensity evolution, the above results combine to yield a theoretical model that can predict the full tangential wind field response to idealized landfalls. An example application to a real world landfall is provided to demonstrate its potential utility for risk applications.
} 

\begin{document}

\maketitle

%
%
%
\statement
A theoretical model that can predict the time-dependent wind field structure of landfalling tropical cyclones (TCs) with limited storm and environmental information is essential for mitigating hazards and allocating public resources. This work provides a first-order prediction of storm size and structure after landfall, which can be combined with existing intensity prediction to form a simple model describing the inland wind field evolution. Results show its potential utility to model real-world inland TC wind fields.
%
%

%
\section{Introduction}\label{intro}

Predicting the inland impacts of a tropical cyclone (TC) depends not only on the evolution of storm intensity (maximum wind speed) but also on the size and structure of the wind field. Empirical models for TC damage that only depend on intensity while neglecting storm size \citep{Mendelsohn2012} significantly underestimate losses \citep{Zhai+Jiang2014}. Storm size is known to vary nearly independently of intensity over the ocean \citep{Frank1977, Merrill1984, Chavas2016,Chavas+Lin2016} and hence predicting the inland impacts requires a model for storm size and structure separate from the intensity. In addition to the direct wind impact, the magnitude and spatial distribution of TC-induced storm surge and heavy rainfall are also strongly dependent on the wind field structure and size \citep{Irish2008, Lu2018}. Larger TCs may also produce more TC tornadoes away from the TC center \citep{Paredes2020}. Accurate estimation of the post-landfall TC wind field, including structure and size, can help prepare for TC hazards and economic losses. 

However, our understanding of the post-landfall evolution of the TC wind field has been limited by insufficient observations and the complexity of landfall processes. The heterogeneity of the inland surface and environmental conditions in the vicinity of the coastline make it difficult to generalize the physics explaining the response of the TC wind field. Recently, \citeauthor{Chen+Chavas2020} (2020, hereafter CC20) simplified landfall as a transient response of a mature axisymmetric TC to instantaneous surface roughening or drying and explained how each surface forcing weakens the storm via different mechanistic pathways using idealized numerical simulation experiments. This work complements idealized 3D landfall experiments that identified important asymmetries in the wind field generated by the onshore flow transition from ocean to land \citep{Hlywiak2021}. \citeauthor{Chen+Chavas2021} (2021, hereafter CC21) generalized this modeling approach to any combination of surface drying and roughening applied simultaneously to test the existing intensification theory of \citep{Emanuel2012} reformulated to predict decay after landfall. They showed that this solution could predict the intensity decay evolution across experiments, and it also compared well with the prevailing empirical intensity decay model. Moreover, they demonstrated that the intensity response to simultaneous drying and roughening could be modeled as the product of the intensity responses to each forcing individually. 

Simple theory for the size and structure of the wind field after landfall has yet to be tested, though. Physical understanding of TC size and structure over the open ocean has advanced significantly in recent decades. Early analytical models proposed an azimuthal wind profile that depend on TC intensity, radius of maximum wind, and the width of the wind maximum \citep{Holland1980}, which have been tested against observations \citep{Shea+Gray1973, Willoughby2004}. Most recently, the theoretical solutions introduced in \cite{Emanuel2004} and \cite{Emanuel+Rotunno2011} can describe the TC low-level wind field in the convection-free outer region and the convective inner region, respectively. For the convective inner region, \cite{Emanuel+Rotunno2011} links the radial variation of the outflow temperature to the radial variation of absolute angular momentum beyond radius of maximum wind via the stratification of the outflow driven by small-scale turbulence. For the convection-free outer region, \cite{Emanuel2004} links the radial gradient of absolute angular momentum to the free troposphere subsidence rate $W_{cool}$, whose value constrained by the heat balance of the free troposphere, via the Ekman dynamics of the boundary layer flow. These two theories are merged together to produce a model for the complete wind profile in \cite{Chavas2015}. The solution takes only a limited number of physical input parameters related to TC intensity, size, latitude, and environmental conditions. This structural model was shown to compare well against observations of the TC wind field and to reproduce the principal modes of wind field variability over the ocean \citep{Chavas+Lin2016}.

TC outer size can vary widely in nature, with size varying more strongly across storms than during the storm life cycle \citep{Chavas2010, Chavas2016}. The size of a given storm tends to depend strongly on the size of its initiating disturbance, with size often growing slowly with time thereafter \citep{Rotunno+Emanuel1987, Martinez2020, XuWang2010}.
On the $f$-plane over an ocean surface, TC size expands towards an equilibrium size that scales with the ratio of the potential intensity to the Coriolis parameter, $V_p/f$ \citep{Wang+Lin+Chavas2022, Frisius2013, Chavas+Emanuel2014}, though slightly different velocity scales have been proposed as well \citep{Khairoutdinov2013, Zhou2014, Zhou2017, Emanuel2022}. Recent work has shown that, on the spherical Earth, TC size is set by the Rhines scale, which depends inversely on the square root of the planetary vorticity gradient $\beta$ \citep{Chavas+Reed2019}. This scaling arises because the TC equilibrium size is much larger than the Rhines scale over the low latitude oceans. Hence, $\beta$ strongly inhibits storms from expanding to their equilibrium size due to Rossby wave radiation \citep{Lu+Chavas2022}. This framework has yet to be considered in the context of landfall, though. Landfall is characterized by a sharp transition to near-zero $V_p$ and thus near-zero equilibrium size \citep{Chen+Chavas2021}. With equilibrium size now much smaller than the Rhines scale, TC size would be expected to shrink towards its equilibrium size after landfall, and its dynamics governed by the length-scale $\frac{V_p}{f}$. We explore this avenue below.
 
Here we examine how existing theory can be used to model the full TC wind field following landfall. This work tests TC size and structure theory against different sets of idealized landfalling storms as in CC21. For structure, we test the theory of \cite{Chavas2015}. For size, we test a simple hypothesis for TC size based on the length scale $\frac{v_m}{f}$, where $v_m$ is used in place of $V_p$ to allow for a transient response to an instantaneous change in $V_p$. We seek to answer the following research questions:
\begin{enumerate}
    \item can the \cite{Chavas2015} wind field model predict the transient response of the azimuthal wind profile to idealized landfall, characterized by instantaneous surface drying and/or roughening?
    
    \item Can the length scale $\frac{v_m}{f}$ predict the transient response of storm outer size?
    
    \item Can the size response to combined drying and roughening be predicted by the product of the responses to each forcing individually, similar to intensity as found in \cite{Chen+Chavas2021}?
    
    \item Can we predict the complete wind field evolution following idealized landfall by combining the structure and size models examined in this work? 

\end{enumerate}

This paper is structured as follows. Section \ref{theory-ch4} reviews the relevant theories. Section \ref{method-ch4} describes model setup and reviews relevant existing theory. Section \ref{results_ch4} presents our results addressing the research questions. Section \ref{summary-ch4} summarizes key results, limitations, and the follow-up work.

\section{Theory}\label{theory-ch4}

\subsection{TC wind structure model}

Absolute angular momentum is widely applied to understand the physics of the TC wind field as it is directly linked to the tangential wind speed \citep{Anthes1974, Montgomery2001, Emanuel2004, Lilly1985}. Within the boundary layer, absolute angular momentum transported inward from some outer radius via radial inflow is gradually lost to surface frictional dissipation. That which remains is gradually converted from planetary to relative angular momentum, thereby generating the tangential wind field of the TC vortex. When reaching the radius of maximum wind speed ($r_m$), air ascends within the convective eyewall and then flows radially outward aloft near the tropopause. Therefore, a theoretical model describing the low-level circulation beyond $r_m$ can be formulated by precisely quantifying how absolute angular momentum changes with radius based on the local dynamics or thermodynamics. Recent work achieves this in two distinct regions: the outer non-convecting region \citeauthor{Emanuel2004} (2004, hereafter E04) and the inner convecting region \citeauthor{Emanuel+Rotunno2011} (2011, hereafter ER11). 

For the convective inner region, small-scale shear-induced turbulence stratifies the outflow to a critical Richardson number $Ri_c$, which for a slantwise neutral vortex translates the stratification of the outflow, $\frac{d T}{dz}$, to a radial increase in $T_0$ with increasing radius beyond $r_m$. The ER11 solution for the radial distribution of $M$ is given by
\begin{equation}\label{ER11theory1}
\left(\frac{M_{ER11}}{M_m}\right)^{2-\frac{C_k}{C_d}}=\frac{2(\frac{r}{r_m})^2}{2-\frac{C_k}{C_d}+(\frac{C_k}{C_d})(\frac{r}{r_m})^2}
\end{equation}
where
\begin{equation}\label{ER11theory2}
M_m=r_m V_m+\frac{1}{2}f {r_m}^2
\end{equation}
is the angular momentum at $r_m$, and $C_k$ and $C_d$ are the exchange coefficients of enthalpy and momentum, respectively. With input metrics ($v_m$, $r_m$) and the value of $C_k$ and $C_d$, Eq.\eqref{ER11theory1}-\eqref{ER11theory2} can generate a complete azimuthal wind profile, though the underlying physics discussed above are only valid for the convecting region beyond $r_m$. 

For the convection-free outer region, the inward flow due to the surface friction torque induces an Ekman suction through the top of the boundary layer from the free troposphere. To satisfy mass continuity, the free troposphere subsidence rate $W_{cool}$ must equal the Ekman suction rate $w_{EK}$,  $-W_{cool}=w_{EK}$, where
\begin{equation}\label{Eq_wek_ch4}
w_{EK}=-\int_{0}^{h} \frac{1}{r}\frac{\partial (ru)}{\partial r} \,\partial z.
\end{equation}
Meanwhile, in this steady-state slab boundary layer with a depth of $h$, the angular momentum $M$ budget is given by
\begin{equation}\label{Eq_BLMbugdet}
hu\frac{\partial M}{\partial r}=-C_d|\textbf{V}|(rV),
\end{equation}
where $\textbf{V}$ is the near surface wind velocity and $V$ is the azimuthal component. Eq.\eqref{Eq_wek_ch4} can be solved by first vertically integrating the slab layer with a depth of $h$ and then radially integrating the layer from $r_0$ to $r$ assuming $u=0$ at $r_0$. Combining the result with Eq.\eqref{Eq_BLMbugdet} to eliminate $hu$, taking $w_{EK} = -W_{cool}$, and approximating $\textbf{V}$ by $V$ yields the solution
\begin{equation}\label{E04theory1}
\frac{\partial{M_{E04}}}{\partial{r}}=\chi \frac{(rV)^2}{{r^2_0}-r^2}
\end{equation}
where $\chi=\frac{2C_d}{W_{cool}}$. $r_0$ is the radius of vanishing wind ($r_{v=0}$), which represents the overall storm outer size. This solution links the radial gradient of $M$ to $W_{cool}$, whose value is constrained by the thermodynamics of free troposphere and can be estimated from the ambient stratification and radiative cooling rate via radiative-subsidence balance. Eq.\eqref{E04theory1} does not have an analytic solution but can be solved numerically to produce a full azimuthal wind profile that extends inwards from $r_0$ to an arbitrary radius. The model takes $r_0$ and $\chi$ as input parameters. 

\citeauthor{Chavas2015} (2015, hereafter referred to C15) mathematically merged the ER11 (Eq.\eqref{ER11theory1}-\eqref{ER11theory2}) and E04 (Eq.\eqref{E04theory1}) solution to produce a model for the complete azimuthal wind profile. This merging yields a unique solution; the process is described in C15. 
Parameters required to solve the solutions are: $V_m$ and $r_m$ for the inner region, $V_a$ and $r_a$ at the merge point connecting the inner and outer region, $r_{fit}$ as a specified radius input, $\chi$ and $f$ for the environmental conditions. Given the environmental parameters $\chi$, $f$, and $C_k/C_d$, one only needs to know two storm parameters -- the intensity $V_m$ and any wind radius (e.g. $r_m$, $r_{34kt}$) -- to specify the model solution. 

C15 is the simplest model that generates a first-order prediction of the full wind field, has been evaluated against real world storms over the ocean, and applied in storm surge risk analysis \citep{Xi2020,Xi2022}. In addition, it performs best among existing wind models in simulating peak storm surge from historical US landfalls \citep{Wang+Lin+Gori2022}. Thus, the C15 wind field model is examined against the simulated wind field response to idealized landfalls in this paper.

\subsection{TC size}
As described in Section 1, TC size on the $f$-plane typically expands towards an equilibrium potential size given approximately by the ratio of the potential intensity to the Coriolis parameter, $\frac{V_p}{f}$. The size of real-world TCs is typically significantly smaller than this length scale \citep{Chavas+Lin2016, Chavas+Reed2019}, though, and instead follows the Rhines scale because the latter is much smaller than $\frac{V_p}{f}$ at low latitudes \citep{Chavas+Reed2019, Lu+Chavas2022}. In contrast, at high latitudes, the potential size is much smaller than the Rhines scale and hence the effect of the Rhines scale becomes negligible, yielding a polar cap regime in aqua-planet experiments and producing a domain filled with TCs analogous to that found on the $f$-plane \citep{Chavas+Reed2019}. Here we propose an analogous regime contrast for landfall: the transition from ocean to land is a transition to a near-zero value of $V_p$ \citep{Chen+Chavas2021} and thus a transition to a regime where $\frac{V_p}{f}$ is suddenly much smaller than the Rhines scale. As with the polar cap regime, the Rhines scale becomes secondary and the TC would be expected to shrink towards its potential size that is near zero. We currently lack an explicit theory for the rate of change of size toward its potential size, though. Instead, we posit that these dynamics ought to depend on $\frac{V_p}{f}$, just as the dynamics of intensity change in nature depends fundamentally on the potential intensity \citep{Tang2012}, including for idealized landfall \citep{Chen+Chavas2021}. Landfall is a transient adjustment between two equilibrium states \citep{Chen+Chavas2020}, and hence size will not scale directly with $\frac{V_p}{f}$ or else the storm would shrink to zero size nearly instantaneously, which clearly does not happen even for an instantaneous transition to land \citep{Chen+Chavas2021}. Thus, we propose the next simplest hypothesis: that size after landfall will scale with $\frac{v_m}{f}$, where $v_m$ is the maximum wind speed itself. Unlike $V_p$, which can change instantly, $v_m$ will change over finite timescale and its response may also be predictable theoretically or empirically \citep{Chen+Chavas2021}. This approach ties storm size to intensity in the transient response in the same manner as it is done at equilibrium via $\frac{V_p}{f}$. We test this hypothesis below.

For the definition of storm outer size, multiple metrics have been applied to define $r_0$ in past work, including $r_{12}$ and $r_{34kt}$, where $r_{12}$ represents the radius of $12 \; ms^{-1}$ wind and $r_{34kt}$ represents the radius of 34 knot azimuthal wind speed. In practice, $r_{34kt}$ is the outermost wind radius that is commonly estimated in operations \citep{Knapp2010, NHC} and can be linked directly to $r_m$ via the structural model described above \citep{Chavas+Knaff2022}. Thus, we choose to focus on $r_{34kt}$ as our outer size metric on practical grounds, as the outer circulation tends to vary coherently \citep{Chavas2015}. 

\section{Methodology}\label{method-ch4}

\subsection{Idealized landfall simulations}

As discussed in CC20 and CC21, spatiotemporal heterogeneity in surface properties are complicated in real-world landfalls, but landfall is fundamentally a transient response to a rapid change in surface wetness and roughness. Here we design simplified landfall experiments in an axisymmetric geometry with a uniform environment and uniform boundary forcing to test the response of a mature TC to modified surface roughness and wetness, individually and in combination. Idealized landfall experiments are performed using the Bryan Cloud Model (CM1v19.8) \citep{Bryan+Fritsch2002} in an axisymmetric geometry with the same setup as CC21. CM1 solves the fully compressible equation of motion in height coordinates on an $f$ plane on a fully staggered Arakawa C-type grid. Model parameters are summarized in Table \ref{geom}. This simple approach neglects all water–radiation and temperature–radiation feedbacks \citep{Cronin+Chavas2019}. The simulation results are robust to varying the choice of model resolutions or mixing lengths \citep{Chen+Chavas2021}. Detailed explanations and discussions about the model setups can be fully referred to CC21. 

We first run a baseline experiment with the above model setup to generate the control experiment (CTRL). The 200-day baseline simulation allows a mature storm to reach a statistical steady-state, from which we identify a stable 15-day period and then define the CTRL as the ensemble mean of five 10-day segments of the baseline experiment from this stable period whose start times are each one day apart. Using ensemble data helps to reduce noise and increases the robustness of the results. During this 10-day evolution, a quasi-stable storm is maintained. Then we perform different types of idealized landfall experiments by restarting each of the five CTRL ensemble members with surface wetness and/or roughness modified. Surface wetness is modified by decreasing surface evaporative fraction $\epsilon$, which reduces the surface latent heat fluxes $F_{LH}$ through the decreased surface mixing ratio fluxes $F_{qv}$ in CM1 (sfcphys.F). Surface roughness is modified by increasing the drag coefficient $C_d$, which alters the surface roughness length $z_0$ and then the friction velocity $u^*$ for the surface log-layer in CM1. Readers are referred to CC20 for full details of the modifications in CM1 experiments. Finally, analogous to the CTRL, the five 10-day segments of each landfall experiment are averaged to reduce noise and yield a single mean response evolution. 

\begin{figure*}[p]
\centerline{\includegraphics[width=0.8\textwidth]{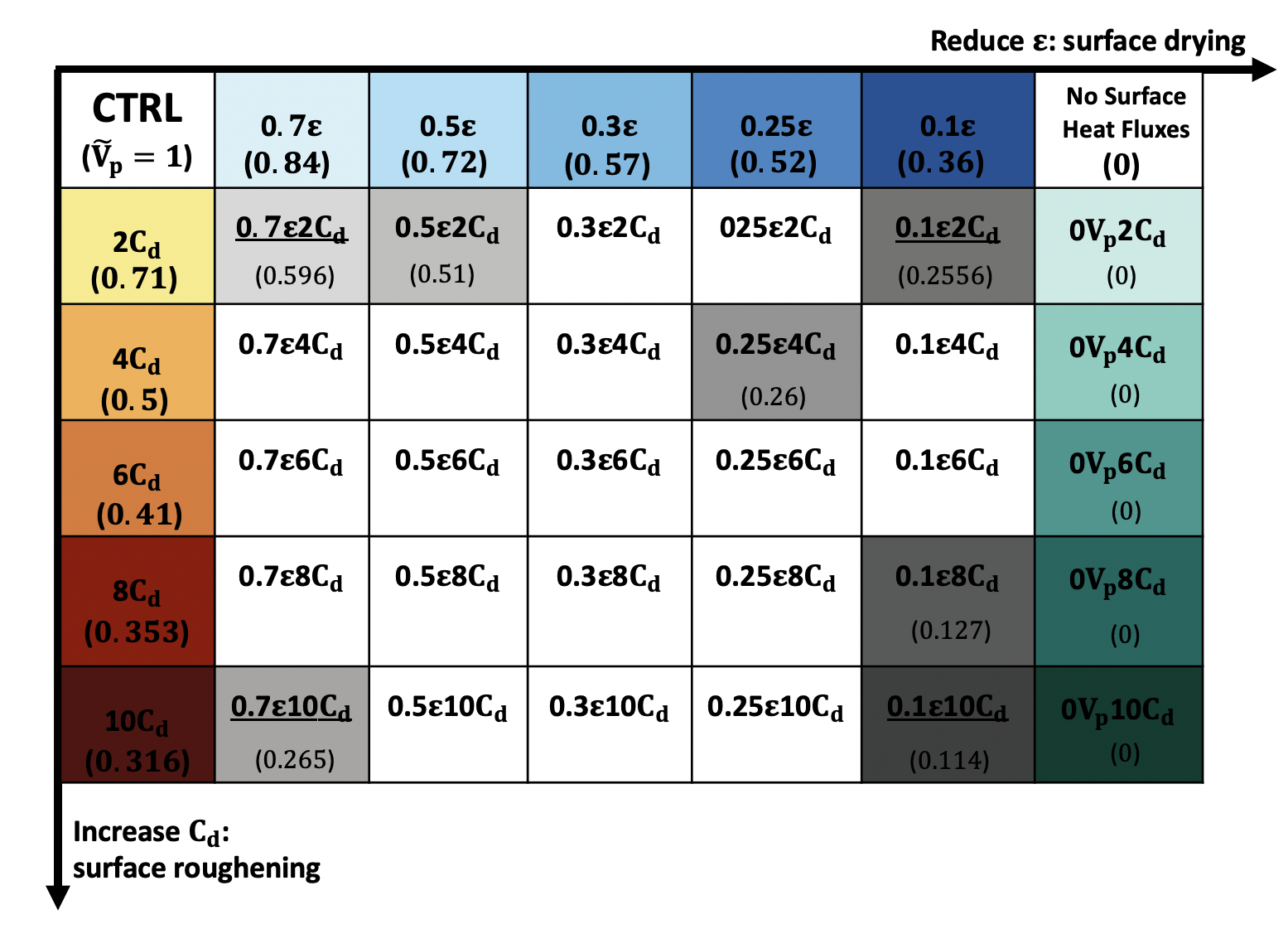}}
\caption{Two-dimensional experimental phase space of surface drying (decreasing $\epsilon$ moving left to right) and surface roughening (increasing $C_d$ moving top to bottom). CTRL is an ocean-like surface with $(C_d,\epsilon) = (0.0015,1)$. Values of the potential intensity response $\tilde V_p$ for CTRL, individual drying or roughening, and representative combined experiments are listed in parentheses; $\tilde V_p$ for any combination of forcing is the product of $\tilde V_p$ for each individual forcing. Representative experiments testing combined forcings are shaded grey and the subset testing the most extreme combinations of each forcing are underlined. Experiment set $0V_pXC_d$, corresponding to the special case where surface heat fluxes are entirely removed ($V_p=0$), are shaded green.}\label{Table_experiments}
\end{figure*}

The design of the idealized landfall experiments is summarized in Fig. \ref{Table_experiments}. It includes roughening-only experiments ($X C_d$, warm color boxes), drying-only experiments ($Y\epsilon$, gradient blue boxes), representative combined experiments ($Y\epsilon X C_d$, grey boxes), and a special set of combined experiments ($0V_pX C_d$, gradient green boxes) where both surface sensible and latent heat fluxes are set to zero while increasing the roughness. $X$ indicates the increase factor applied to surface drag coefficient $C_d$, while $Y$ is the reduction factor applied to the surface evaporative fraction $\epsilon$. The modification in $C_d$ or/and $\epsilon$ systematically weakens the CTRL storm, which can be understood via the response of potential intensity, $\tilde V_p=\frac{V_{p,EXP}}{V_{p,CTRL}}$ (Eq.4-6 in CC21; Fig.\ref{Table_experiments}). We select two subsets of combined-forcing experiments for deeper analysis: $0.7\epsilon2C_d$, $0.7\epsilon10C_d$, $0.1\epsilon2C_d$, and $0.1\epsilon10C_d$ (underlined in Fig.\ref{Table_experiments}) represent the extreme combinations where each forcing takes its highest or lowest nonzero magnitude. $0.5\epsilon2C_d$, $0.25\epsilon4C_d$, and $0.1\epsilon8C_d$ represent cases where the individual forcing in a combined experiment yields similar contributions to $\tilde V_p$. Finally, we have a special set of combined experiments, $0V_pXC_d$, in which $V_p$ is fully reduced to zero for a range of magnitudes of roughening, which is achieved by setting both surface sensible and latent heat fluxes to zero while increasing the roughness by a factor of $X$. 

\subsection{Testing theory against simulations}

\begin{figure*}[p]
\centerline{\includegraphics[width=\textwidth]{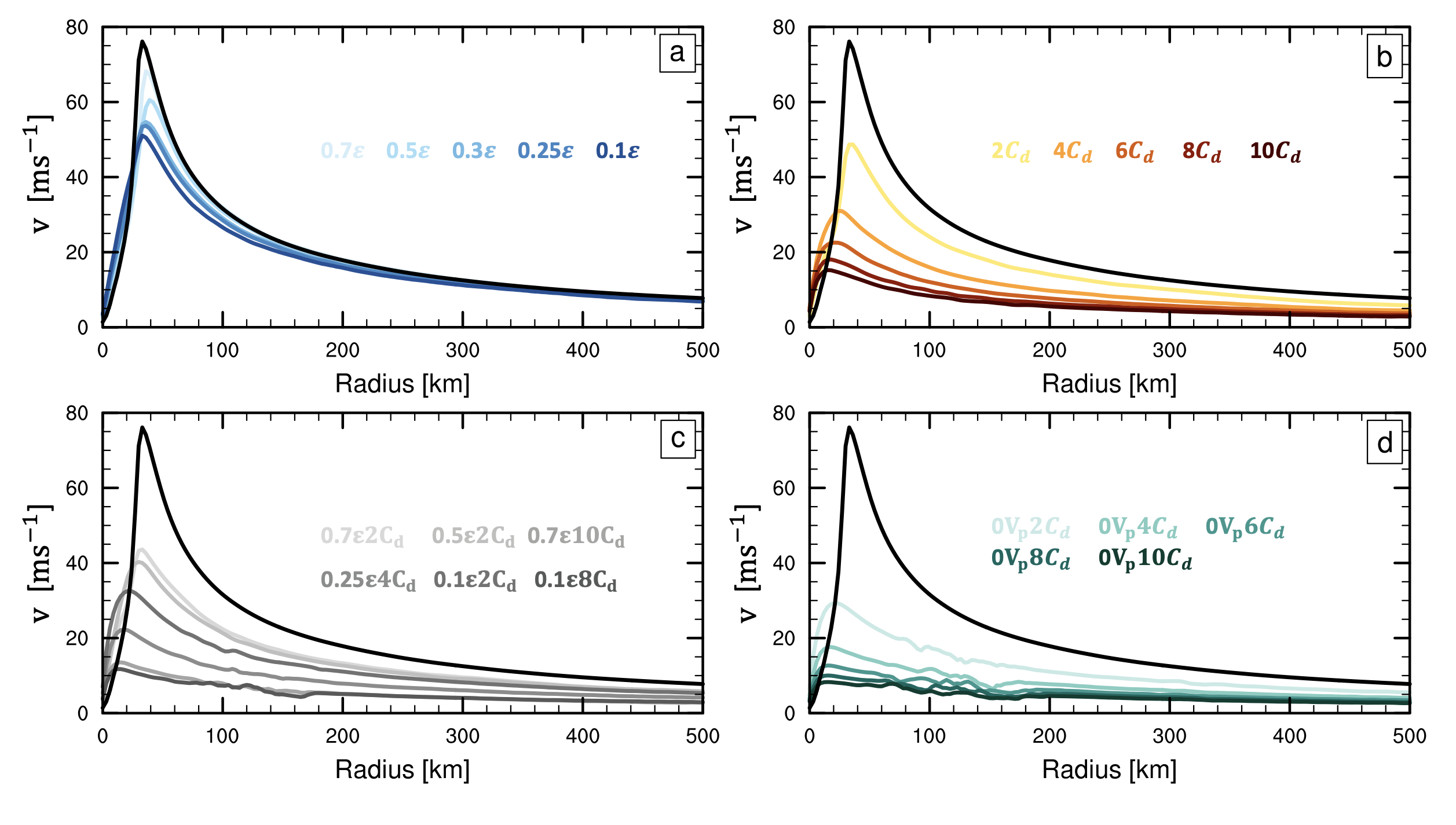}}
\caption{The simulated 10-m wind field in response to different magnitudes of each type of idealized landfall (color lines) at $\tau=24\;h$. The initial CTRL wind field is shown by the thick black line in each plot.}\label{F2}
\end{figure*}

The near-surface wind field drives inland TC hazards. Thus, the evolution of the 10-m wind field in different types of idealized landfall experiments is compared to the C15 model prediction. We focus principally on the first 24 hours of the evolution, during which the wind field response to each landfall type is the strongest while the circulation remains sufficiently well defined to easily identify $r_{34kt}$ (Fig.\ref{F2}). A stable gale wind radius is often no longer observable 24 hours after TC landfall in the real world \citep{Jing+lin2019}. We compare model versus simulation at $\tau=1,6,12,24\;h$ and display results whenever a value of $r_{34kt}$ is still discernible. 

For the C15 prediction, the simulated $v_m(\tau)$ and $r_{34kt}(\tau)$ response of each experiment are used to fit the structural model, where $\tau$ denotes the time since the start of a given forcing experiment. We hold $C_k$ fixed at its CTRL value ($C_k=0.0015$) and set $C_d$ to be its modified value for each experiment. Since latitude typically varies minimally for a landfalling TC over a 24-hour period, the Coriolis parameter $f$ is held fixed at its CTRL value ($5\times 10^{-5} s^{-1}$. The radiative-subsidence rate $W_{cool}$ is set to $0.002\;ms^{-1}$, which is the median of the best-fit value for observed storms \citep{Chavas2015}. The wind field solution is not very sensitive to $W_{cool}$ except for at large radii well beyond $r_{34kt}$ (Supplementary Figure 1). 
For TC size, we test the simple hypothesis
\begin{equation}\label{Eq_sizescale2}
r_{34kt}(\tau)\sim \frac{v_m(\tau)}{f}
\end{equation}
that was motivated in the previous section. Since we hold $f$ constant, this hypothesis simplifies to
\begin{equation}\label{Eq_sizescale_theory}
\tilde{r}_{34kt}(\tau)\sim \tilde{v}_{m}(\tau)
\end{equation}
As in CC21, we analyze results as responses relative to initial state. Hence, the intensity and size in each experiment are normalized by the CTRL value, i.e. $\tilde v_m=\frac{v_{m,EXP}}{v_{m,CTRL}}$ and $\tilde r_{34kt}=\frac{r_{34kt,EXP}}{r_{34kt,CTRL}}$. That is, we examine whether the simulated transient response of intensity can predict the transient response of storm outer size during the decay evolution. 

\section{Results}\label{results_ch4}

\subsection{TC structure}

\begin{figure*}[p]
\centerline{\includegraphics[width=0.8\textwidth]{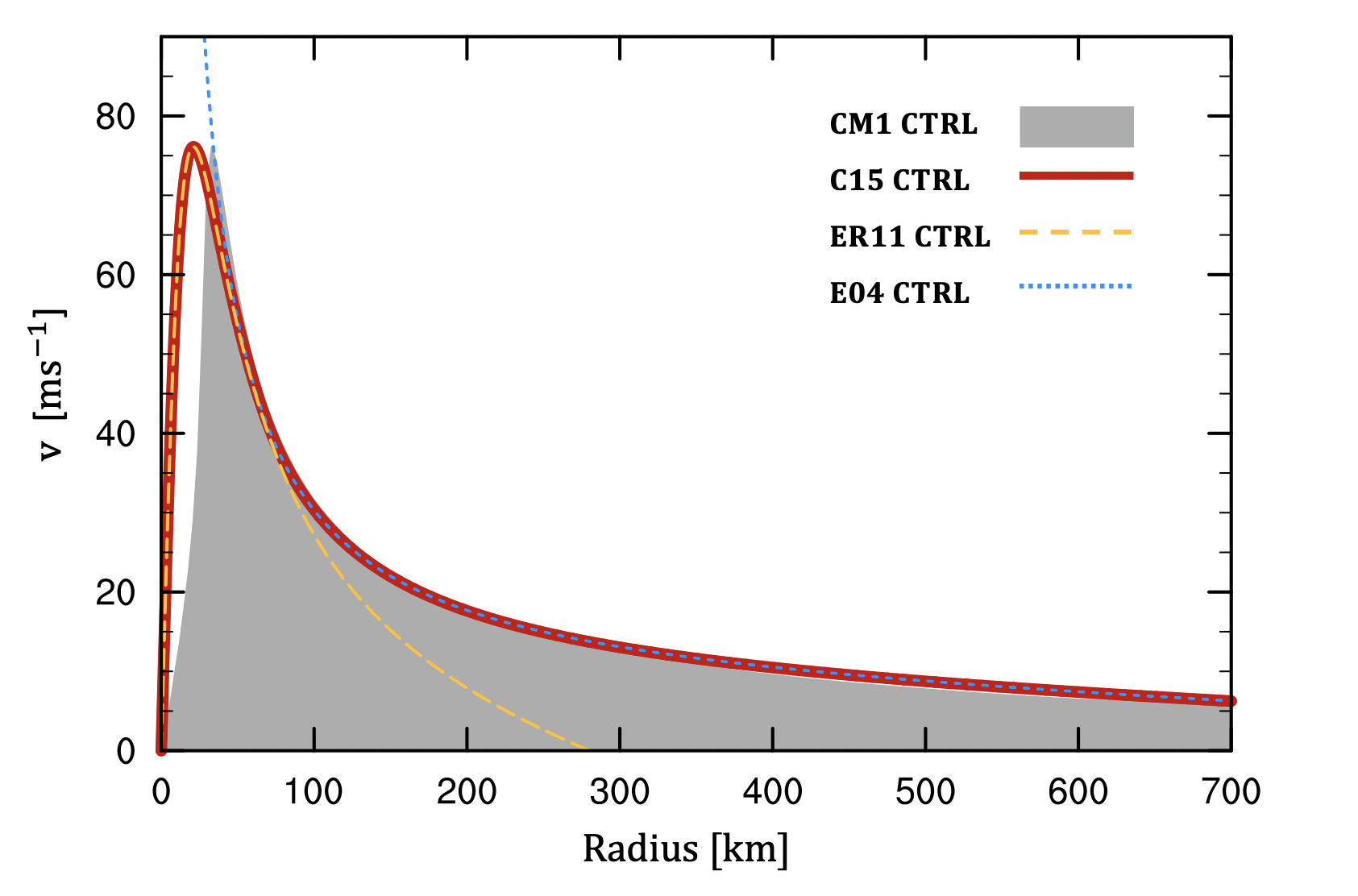}}
\caption{The 10-m wind field of the CM1 simulated steady-state, mature storm (shaded CM1 CTRL), the ER11-predicted wind profile (yellow dash), the E04-predicted wind profile (blue dash), and the C15-predicted wind profile (red line). $f=5x10^{-5} s^{-1}$, $\chi=1.5$ where $C_d=0.0015$ and $W_{cool}=0.002 ms^{-1}$. $v_m=76.1 ms^{-1}$, $r_m=33\;km$, and $r_{34kt}=202\;km$ in the CM1 CTRL simulation, respectively.  }\label{F1theoWind}
\end{figure*}

We first examine to what extent the C15 model can reproduce the response of the storm wind field structure in our idealized landfall experiments, with intensity and size taken as their simulated values. Fig.\ref{F1theoWind} shows the C15 model and its inner and outer component models against a Cloud Model (CM1) simulated low-level (10-m) wind field of a mature storm (CM1 CTRL). The model inputs are the CM1 CTRL values of $v_m$ and $r_{34kt}$. $f$ is set to 0.00005 $s^{-1}$, $W_{cool}=0.002\; ms^{-1}$, and $C_d$ is 0.0015. In this example, the model does very well in predicting the simulated wind profile for nearly the entire circulation beyond $r_m$, but it underestimates $r_m$.

\begin{figure*}[p]
\centerline{\includegraphics[width=0.9\textwidth]{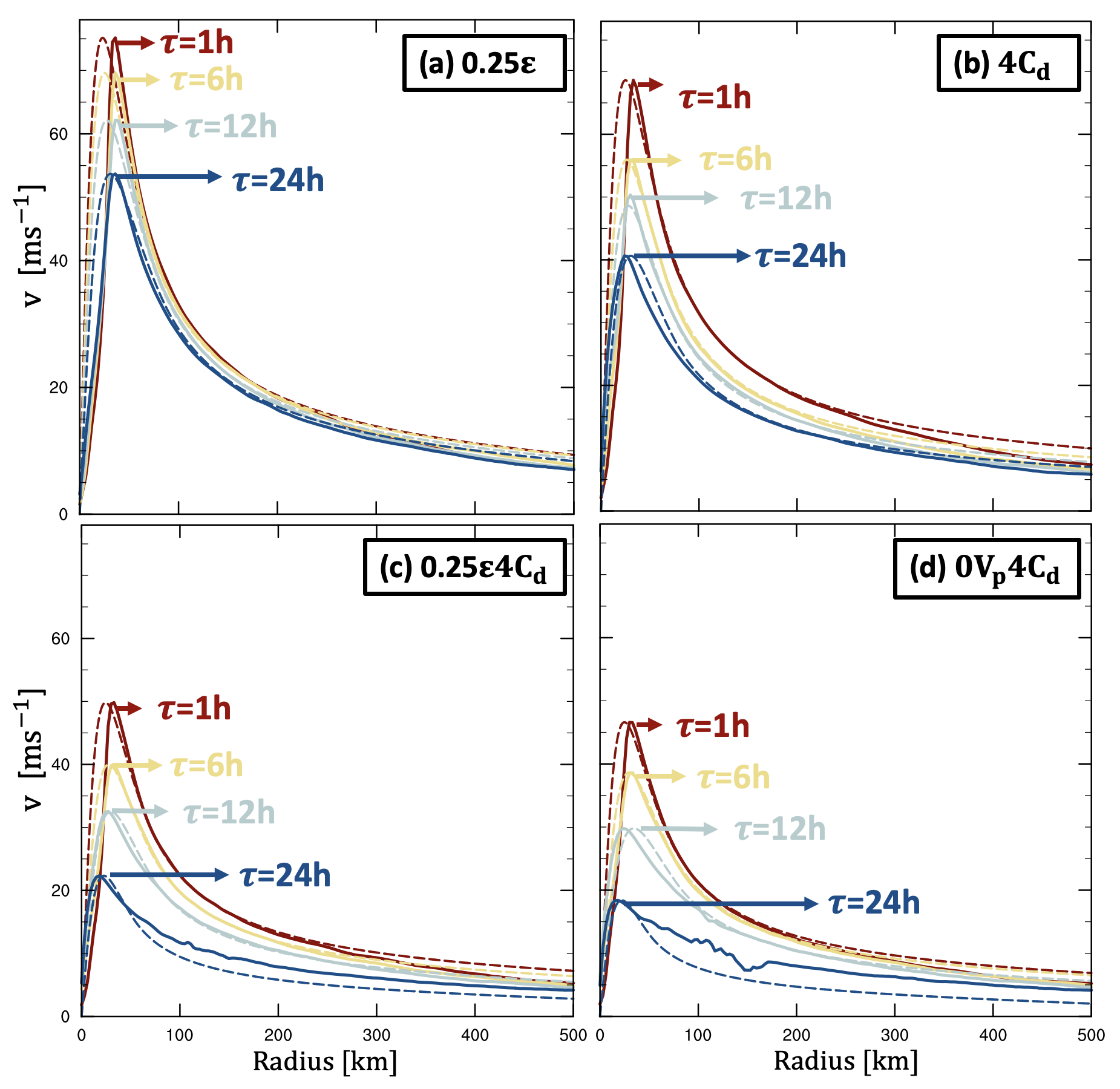}}
\caption{The 10-m wind field of the representative landfall experiments (a) $0.25\epsilon$, (b) $4C_d$, (c) $0.25\epsilon4C_d$ and (d) $0V_p4C_d$ at $\tau=1, 6, 12, 24\; h$. Colored curves are the simulated wind field (solid) and C15-predicted wind field (dash).}\label{F3C15}
\end{figure*}

\begin{figure*}[p]
\centerline{\includegraphics[width=0.9\textwidth]{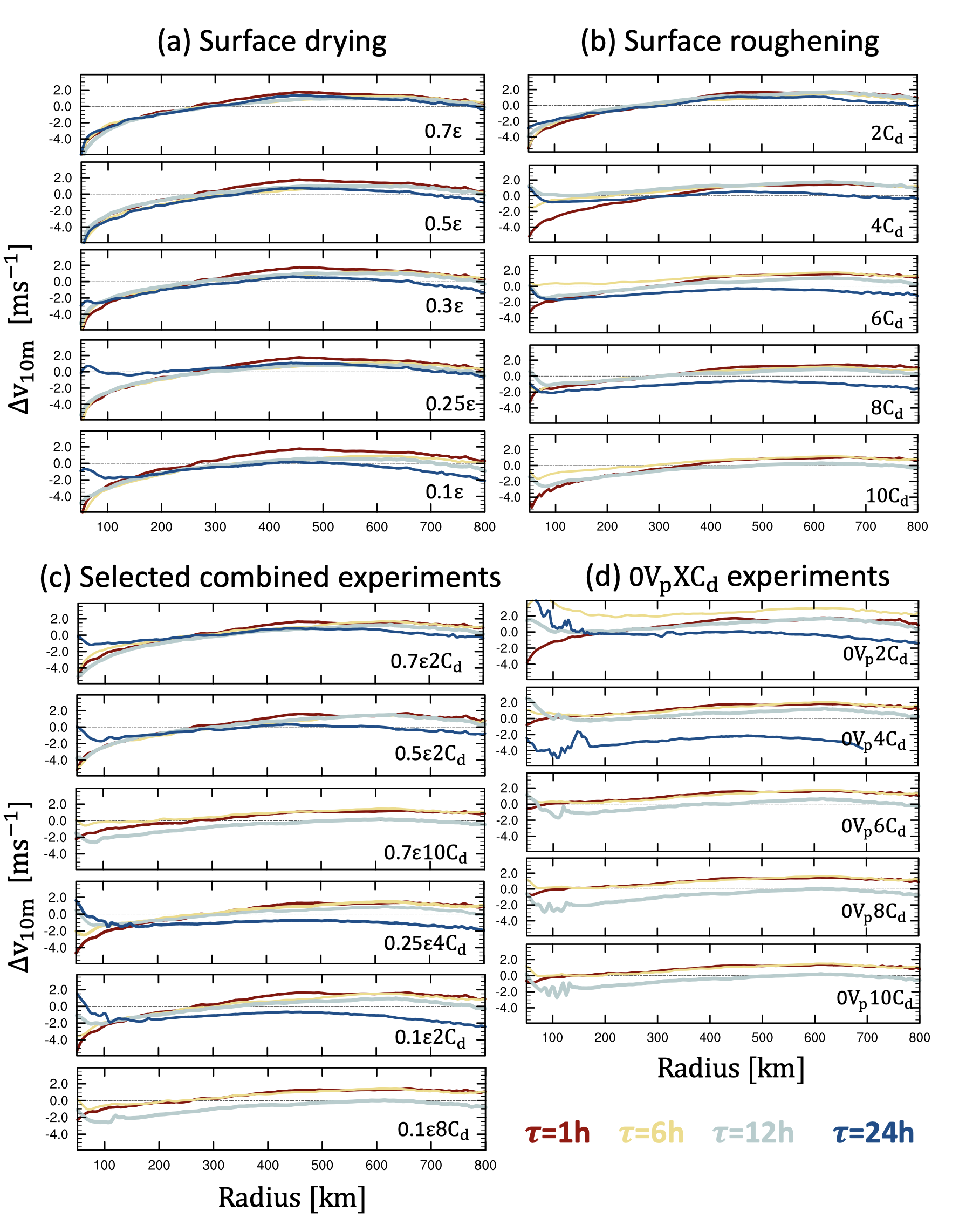}}
\caption{The wind field difference between C15-prediction and the idealized landfall simulation from $r=50$ to $800\;km$ across all the experiments of (a) surface drying experiments, (b) surface roughening experiments, (c) combined experiments and (d) $0V_pXC_d$ experiments at $\tau=1, 6, 12, 24\; h$. Each $\tau$ is indicated by the same color as in Fig.\ref{F3C15}.}\label{DiffC15}
\end{figure*}

Comparisons of the C15 prediction fit to ($v_m$, $r_{34kt}$) and the model simulation for representative idealized landfall experiments $0.25\epsilon$, $4C_d$, $0.25\epsilon4C_d$, and $0V_p4C_d$ at $\tau=1,6,12,24\;h$ are shown in Fig.\ref{F3C15}. The wind field difference between C15 prediction and model simulation across all the experiments are shown in Fig.\ref{DiffC15}.

For surface drying only  (Fig.\ref{F3C15}a), C15 does well in reproducing much of the wind field beyond $2r_m$ out to large radii ($r=800\;km$) during the slow decay. The model generally underestimates the wind speed within the convective inner region ($r\le200\;km$) due to the low bias in $r_m$ that persists and acts to shift the peak wind region slightly inward of the simulation. At $\tau=24\;h$, the low bias near $r_m$ becomes smaller for stronger drying ($0.3\epsilon$, $0.25\epsilon$, $0.1\epsilon$ in Fig.\ref{F3C15}a and Fig.\ref{DiffC15}a). Beyond $r_{34kt}$, the C15 prediction bias is less than $2 ms^{-1}$ across all the drying experiments (Fig.\ref{F3C15}a). 

For experiments that include surface roughening (i.e. roughening-only, combined, and $0V_pXC_d$), the C15 prediction performs well in reproducing the wind field evolution at all radii beyond $r_m$ (Fig.\ref{F3C15}b-d). In contrast to the drying-only experiments, the bias near $r_m$ at the initial timestep decreases with time (Fig.\ref{DiffC15}b-c). The bias is generally less than $2 ms^{-1}$ outside $r_{34kt}$ across all the roughening and combined experiments (Fig.\ref{DiffC15}b-c) similar to drying-only experiments. It is noteworthy that in combined or $0V_pXC_d$ experiments, when $v_m$ approaches 34 knots at $\tau=24\;h$, $r_{34kt}$ locates within the convecting inner-core region where the model is more biased, C15 underestimates the outer wind field (Fig.\ref{F3C15}d and Fig.\ref{DiffC15}d).


\subsection{TC size}

\begin{figure*}[p]
\centerline{\includegraphics[width=\textwidth]{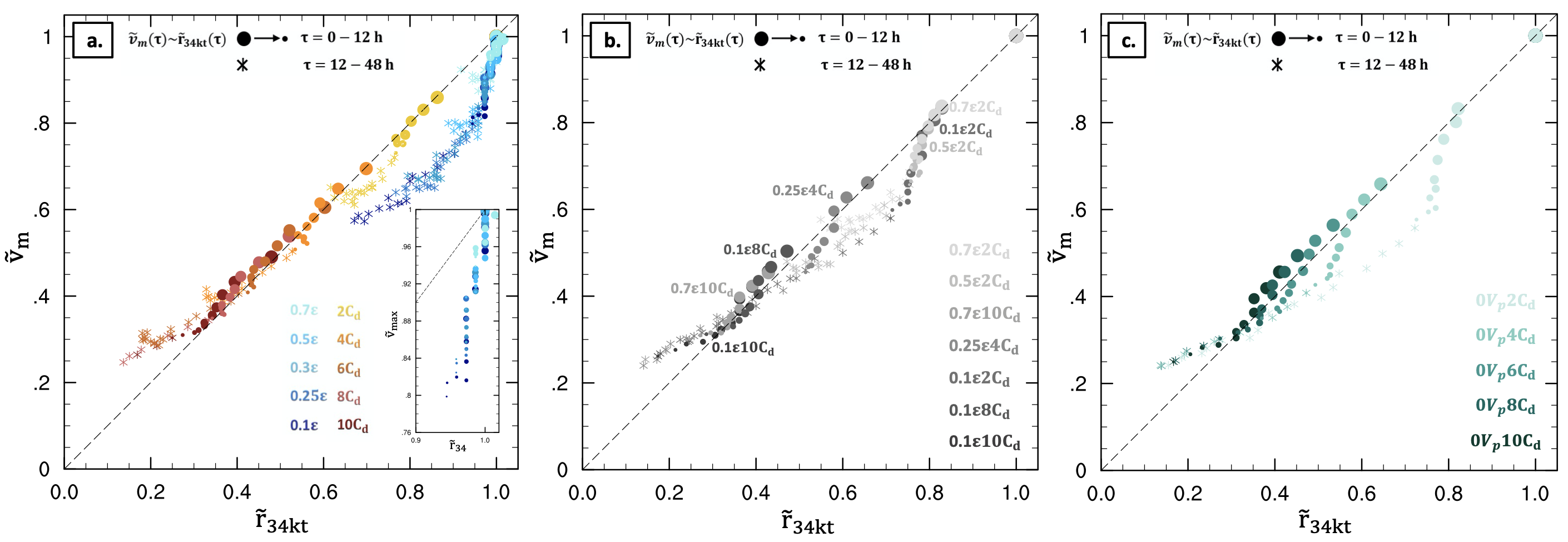}}
\caption{The relationship between $\tilde v_m(\tau)$ and $\tilde r_{34kt}(\tau)$ during a 48-h evolution (asterisk is marked by every 2 $h$) for (a) individual forcing experiments, (b) combined experiments and (c) $0V_pXC_d$ experiments. The subplot in (a) emphasizes on the relationship between $\tilde v_m(\tau)$ and $\tilde r_{34kt}(\tau)$ during the first 12-h period for all surface drying experiments. For $0V_pXC_d$ experiments with stronger surface roughening ($0V_p8C_d$, $0V_p10C_d$), the relationship no longer exits after $\tau=24\;h$ due to the rapid decay.} \label{F4sizescale}
\end{figure*}

We next examine to what extent the transient response of size $\tilde r_{34kt}(\tau)$ scales with the transient response of intensity $\tilde v_m(\tau)$ in individual-forcing experiments (Fig.\ref{F4sizescale}a), combined-forcing experiments (Fig.\ref{F4sizescale}b), and $0V_pXC_d$ experiments (Fig.\ref{F4sizescale}c) throughout the 48-h evolution. For cases with strong surface roughening ($8C_d$, $10C_d$, $0.7\epsilon10C_d$, $0.1\epsilon8C_d$, $0.1\epsilon10C_d$ and $0V_p8C_d$, $0V_p10C_d$), simulated intensity quickly decreases below 34 knots after $\tau=12\;h$ and thus their corresponding long-term $\tilde {r}_{34kt}(\tau)\sim \tilde v_m(\tau)$ relationship is not shown in Fig.\ref{F4sizescale}.

For roughening experiments (Fig.\ref{F4sizescale}a, red colors), $\tilde {r}_{34kt}$ scales very closely with $\tilde{v}_{m}$ throughout the 48 hour period, and it does so consistently across all experiments. For drying experiments (Fig.\ref{F4sizescale}a, blue colors), $\tilde {r}_{34kt}$ scales reasonably closely with $\tilde{v}_{m}$ by the end of the 48 hour period. However, during the first 12h, the storm principally weakens while its outer size slowly shrinks (Fig.\ref{F4sizescale}a subplot) as found in CC20, and then the size subsequently begins to shrink steadily with the decreasing intensity. During the last 12h, the size continues to gradually shrink after intensity becomes steady, especially in $0.7\epsilon$, $0.5\epsilon$, and $0.3\epsilon$ experiments. Thus, $\tilde r_{34kt}$ scales more closely with $\tilde v_m$ at the end of the evolution, with $\tilde r_{34kt}$ ending up about 10-15\% higher than $\tilde v_m$ across all the drying experiments. The trajectories through ($\tilde{v}_{m},\tilde {r}_{34kt}$) space is consistent across all experiments similar to the roughening experiments. 

For combined experiments and $0V_pXC_d$ experiments, storm size response $\tilde {r}_{34kt}(\tau)$ also generally scales with the intensity response $\tilde v_m(\tau)$ throughout the evolution (Fig.\ref{F4sizescale}b-c). However, $\tilde {r}_{34kt}(\tau)$ exhibits both characteristics of surface drying and roughening experiments. During the initial period ($\tau=0-6\;h$), $\tilde {r}_{34kt}(\tau)$ decreases sharply and nearly linearly with $\tilde v_m(\tau)$ across all the experiments. For $\tau=6-12\;h$, for experiments with weaker roughening ($0.7\epsilon2C_d$, $0.5\epsilon2C_d$, $0.1\epsilon2C_d$ in Fig.\ref{F4sizescale}b and $0V_p2C_d$ in Fig.\ref{F4sizescale}c), $\tilde {r}_{34kt}(\tau)$ remains relatively steady before decreasing again after $\tau=12\;h$ to follow a trajectory similar to roughening out to $\tau=48\;h$. 

\begin{figure*}[p]
\centerline{\includegraphics[width=\textwidth]{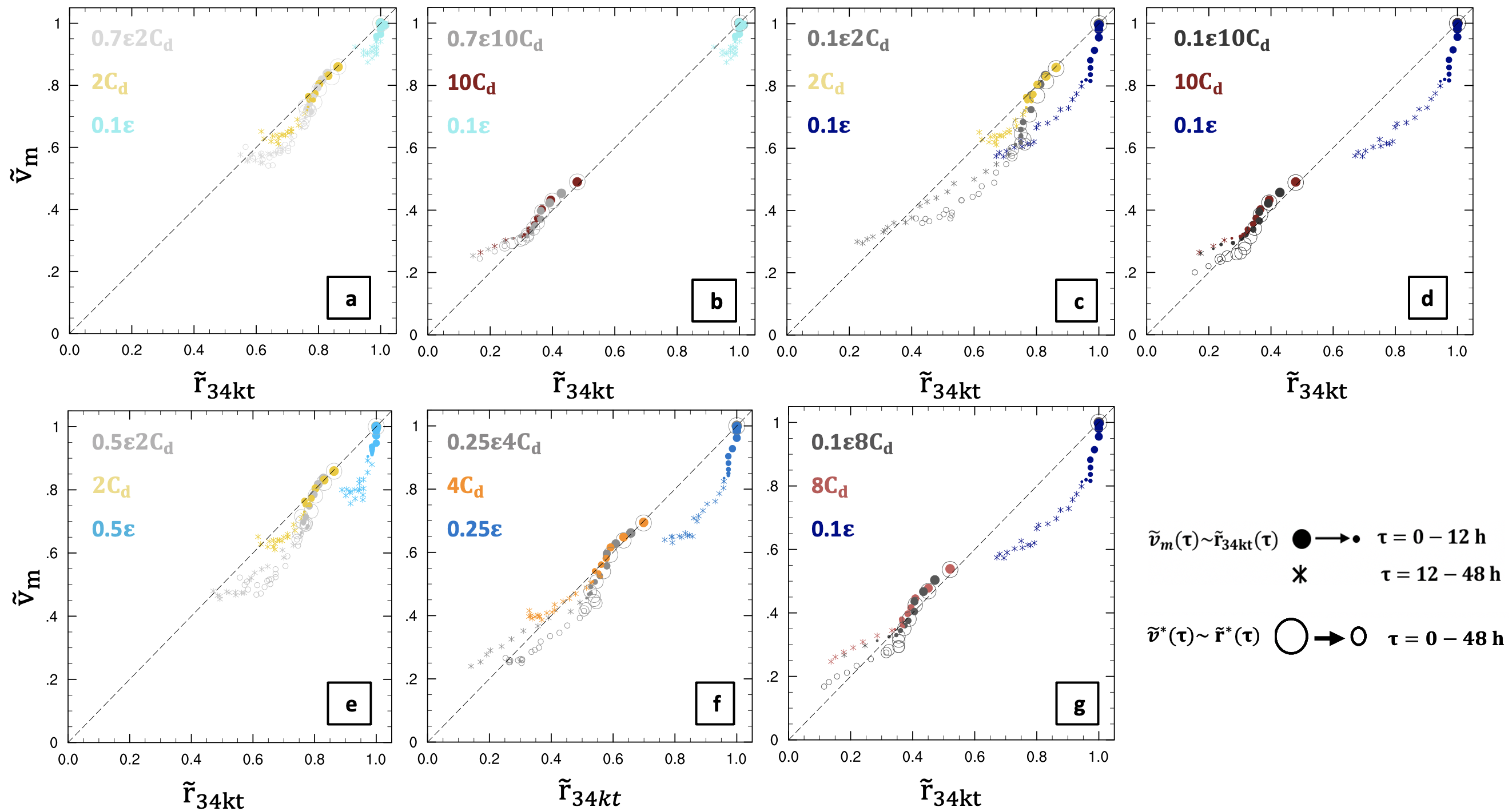}}
\caption{The relationship between simulated $\tilde v_m(\tau)$ and $\tilde r_{34kt}(\tau)$ for combined experiments and their deconstructed individual forcing experiments (asterisk is marked by every 2 $h$), and the predicted $\tilde v^*(\tau)$ and $\tilde r^*(\tau)$ relationship (Eq.\eqref{Eq_vmdeconstruct}-\eqref{Eq_v_r34hypo}) of each combined experiment (decreasing-size circles plotted by every 2 $h$). Here (a) $0.7\epsilon2C_d$, (b) $0.7\epsilon10C_d$, (c) $0.1\epsilon2C_d$, and (d) $0.1\epsilon10C_d$ are representative combined experiments where each individual forcing takes its highest or lowest non-zero magnitude. (e) $0.5\epsilon2C_d$, (f) $0.25\epsilon4C_d$, and (g) $0.1\epsilon8C_d$ are representative combined experiments where the magnitude of individual forcing in a combined experiment yields similar contribution to the intensity response.} \label{F5sizescale}
\end{figure*}

Next, we deconstruct the response of the storm size for combined experiments to explore whether their $\tilde r_{34kt}$ can be predicted via deconstructed physical processes caused by individual surface roughening and drying. As found in CC21, the complete time-dependent response of storm intensity to simultaneous surface roughening and drying, $\tilde v_{m,C_d\epsilon}(\tau)$, can be predicted by the product of the individual response, $v^{*}(\tau)$, as
\begin{equation}\label{Eq_vmdeconstruct}
 \tilde v_{m,C_d\epsilon}(\tau)\approx \tilde v^{*}(\tau)=\tilde v_{m,C_d}(\tau)\tilde v_{m,\epsilon}(\tau),
\end{equation}
Given that we find that the $\tilde {r}_{34kt}(\tau)$ scales reasonably well with $\tilde v_m(\tau)$, we propose a similar hypothesis here for the size response in combined experiments as
\begin{equation}\label{Eq_r34deconstruct}
 \tilde r_{34kt,C_d\epsilon}(\tau)\approx \tilde r^{*}(\tau)=\tilde r_{34kt,C_d}(\tau)\tilde r_{34kt,\epsilon}(\tau),
\end{equation}
and assume that
\begin{equation}\label{Eq_v_r34hypo}
 \tilde v^{*}(\tau) \sim \tilde r^{*}(\tau).
\end{equation}
The hypotheses made in Eq. \eqref{Eq_r34deconstruct} and \eqref{Eq_v_r34hypo} are tested against our representative combined simulations (Fig.\ref{F5sizescale}). 

First we compare the estimated responses $\tilde r^{*}(\tau)$ and $\tilde v^{*}(\tau)$ to the corresponding simulated responses, $\tilde {r}_{34kt}(\tau)$ and $\tilde v_m(\tau)$, of the combined experiments (Fig.\ref{F5sizescale}, grey markers). Overall, the estimated size-intensity relationship $\tilde r^{*}(\tau) \sim \tilde v^{*}(\tau)$ follows the simulated relationship $\tilde {r}_{34kt}(\tau)\sim \tilde v_m(\tau)$ closely throughout the evolution across all the combined experiments. 

Finally, we identify the dominant forcing for driving the responses in $\tilde v_m(\tau)$ and $\tilde {r}_{34kt}(\tau)$ in our combined experiments. The outer size responses in the combined experiment and the analogous roughening-only experiment with identical $C_d$ are similar during the initial 12 h, during which size and intensity change most strongly, before deviating thereafter (Fig.\ref{F5sizescale}, grey and warm color markers). This consistency suggests that the size response is primarily dominated by surface roughening, regardless of its magnitude or the concurrent magnitude of drying. Surface drying imposes an impact on the storm size largely during the later period ($\tau>12\;h$) after the storm has already weakened considerably.

To summarize, ${r}_{34kt}(\tau)$ scales quite closely with ${v_m(\tau)}$ for mature storms in response to idealized landfall, especially for a rougher land surface. This finding provides some evidence for our hypothesis that the equilibrium size length-scale on the $f$-plane becomes important for the dynamics of the transition to land; testing the role of $\beta$ lies beyond the scope of this work. The above results suggest that a viable simple prediction of the storm outer size for idealized experiments with any combined forcing can be made if given the estimation of intensity response to corresponding surface modifications (Eq.\eqref{Eq_v_r34hypo}).

All our results taken together suggest the potential to predict the complete wind field evolution to idealized landfall if given the intensity response. We explore this avenue next.

\subsection{A model for the wind field in idealized landfalls}\label{CC21+C15}

Finally, we combine the findings for modeling structure and size in this work to predict the response of the near-surface wind field and compare against our subsets of combined-forcing landfall experiments. 

\begin{figure*}[p]
\centerline{\includegraphics[width=\textwidth]{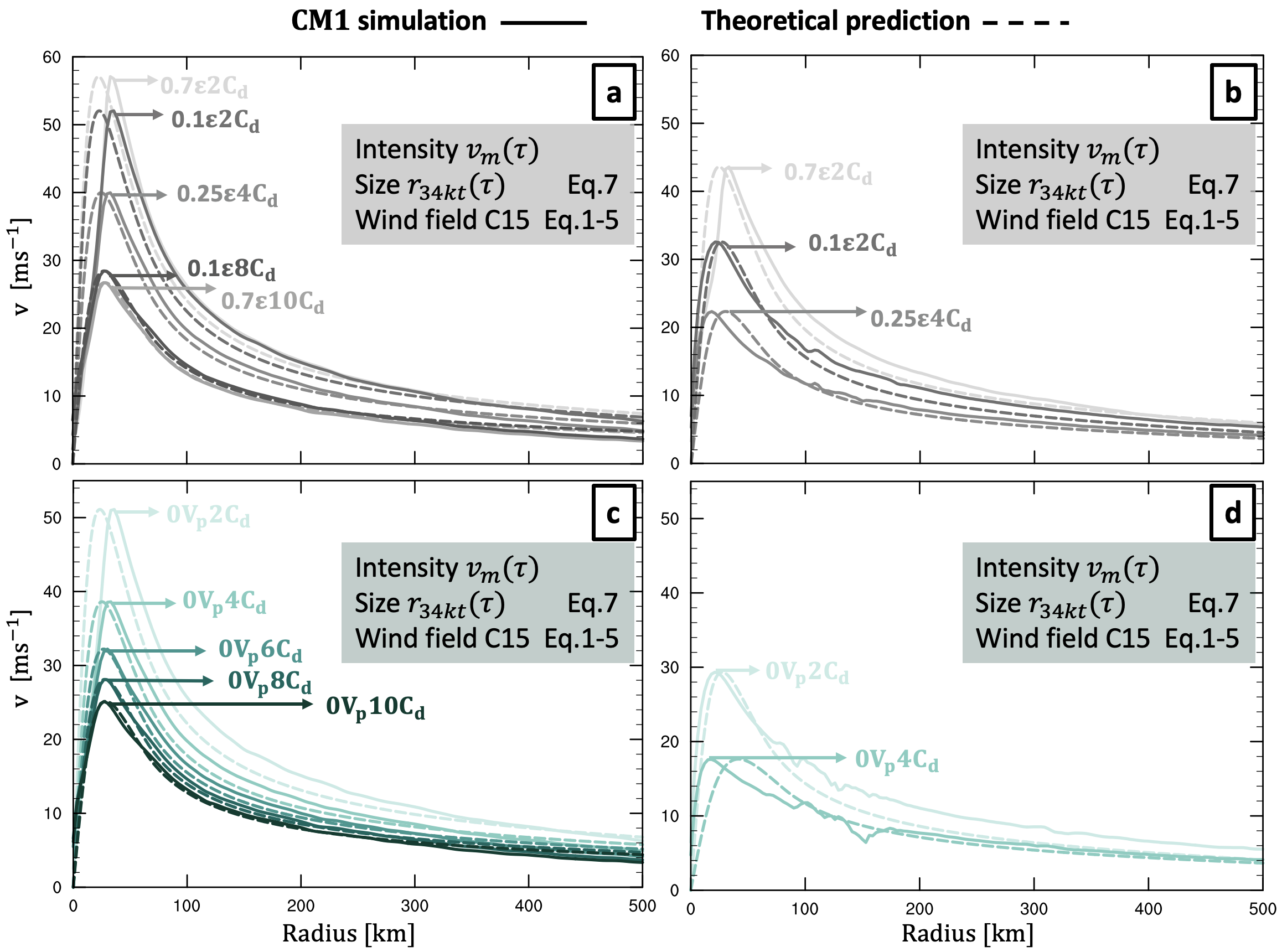}}
\caption{Simulated Wind field of representative landfall experiments (solid) and the corresponding theoretical prediction (dash) at (a)(c) $\tau=6\;h$ and (b)(d) $\tau=24\;h$. The theoretical prediction applies the simulated intensity response $v_m(\tau)$ and the corresponding size prediction $r_{34kt}(\tau)$ (Eq.\eqref{Eq_sizescale_theory}) in the C15 model. Experiments with stronger surface modifications decay too fast to produce the input $v_m$ and $r_{34kt}$ for C15 model at $\tau=24\;h$, thus those experiments ($0.1\epsilon8C_d$, $0.7\epsilon10C_d$ and $0V_p6C_d$, $0V_p8C_d$, $0V_p10C_d$) are not shown in (b)(d).} \label{F7C15CC21}
\end{figure*}

We model the wind field using the C15 model with inputs as described above, which requires the temporal evolution of intensity and size. Here we use the simulated intensity evolution $v_{m}(\tau)$. We then predict the outer size evolution $\tilde r_{34kt}(\tau)$ by assuming it scales directly with $\tilde v_{m}(\tau)$ as in Eq.\eqref{Eq_sizescale_theory}. Note that the intensity evolution could also be predicted via a statistical model such as \citep{Jing+lin2019} or theory-based model such as that proposed in CC21, though we do not do so here in order to focus on the representation of the wind field. Knowing the initial intensity and size, $v_{m,0}$ and $r_{34kt,0}$ at $\tau=0\;h$ (i.e., just prior to idealized landfall), one can produce $r_{34kt}(\tau)$ via $v_{m}(\tau)$. As an example, the theoretical prediction of the wind profile at $\tau=6\;h$ and $24\;h$ is compared to the simulated profile for each experiment in Fig.\ref{F7C15CC21}, where $0.1\epsilon8C_d$, $0.7\epsilon10C_d$ and $0V_p6C_d$, $0V_p8C_d$, $0V_p10C_d$ are excluded at $\tau=24\;h$ since their corresponding intensity quickly decreases below 34 knots (Fig.\ref{F7C15CC21}b and d). 

Overall, the wind field model prediction performs reasonably well in capturing the simulated wind field response across all experiments. The prediction of $r_m$ itself is imperfect, especially for experiments where surface is less roughened (Fig.\ref{F7C15CC21}), since the model begins with the low bias in $r_m$ from CTRL and is unable to describe the wind field inside $r_m$. 
For weak roughening (Fig.\ref{F7C15CC21}b and d), as $v_m$ decreases approaching 34 knots in later stages, the model more strongly underestimates the simulated size of the inner region wind field. For strong roughening, though, (Fig.\ref{F7C15CC21}a and c) this inner-core bias tends to decrease with time.

Note that we also tested this simple wind field model using the intensity model of CC21 $v^{*}_{th}(\tau)$ introduced in CC21 (Eqs.14-15 therein) and its corresponding size estimation $r^{*}_{th}$ via Eq.\eqref{Eq_v_r34hypo} (Supplementary Figure 2). This approach also works reasonably well, though it still requires specification of the boundary layer depth parameter, which is poorly-constrained as described in CC21. Hence we have focused here on taking intensity as known, as alternative intensity models may be equally viable in practice.

To summarize, given the TC intensity and $r_{34kt}$ prior to landfall and knowledge of the idealized land surface conditions, one can predict the first-order post-landfall wind field evolution. In contrast to the intensity decay model of CC21, which depends on a poorly-understood boundary layer height parameter, the size and structure results presented here do not depend on any free parameters and hence are expected to apply generally. This simple model may serve as a foundation for a model to predict the wind field response to landfall that further incorporates the many additional complexities associated with real-world landfalls.

\section{Summary and Discussion}\label{summary-ch4}

This work proposes a simple theory-based model for the response of the tropical cyclone wind field to idealized landfalls and tests it against numerical simulation experiments. The model combines an existing physics-based model for the wind field and a simple model for storm outer size $r_{34kt}(\tau)$ that assumes it follows the response of maximum wind speed $v_m(\tau)$. Combining these results with TC intensity prediction, this work provides a theoretical model for inland TC wind field. Key findings are as follows:

\begin{itemize}

\item Given simulated $v_m$ and $r_{34kt}$, the C15 wind field model (Eq.\eqref{ER11theory1}-\eqref{E04theory1}) generally reproduces the response of the wind field beyond $r_m$ to idealized landfalls over the first 24-h, which is the period of most significant weakening. For the convecting inner-core region near $r_m$, the C15 model is not able to precisely predict the $r_m$, though this is due in part to a bias in the initial profile itself. For the convection-free outer region, the C15 prediction generally reproduces the wind field response to various forcings with minimal bias over much of the circulation beyond $r=50\;km$. 
  
\item The landfall response of storm size $r_{34kt}(\tau)$ is found to scale closely with that of storm intensity $v_m(\tau)$ (Eq.\eqref{Eq_sizescale_theory}), particularly in the presence of relatively strong roughening. This finding aligns with the hypothesis that the equilibrium storm size length-scale, $V_p/f$, becomes important in the dynamics of the transition to land where $V_p$ becomes near zero. 

\item The storm size response to combined drying and roughening can be deconstructed as the product of the responses to each individual forcing (Eq.\eqref{Eq_vmdeconstruct}-\eqref{Eq_v_r34hypo}), similar to intensity as found in CC21. Surface roughening imposes a strong and rapid initial response and hence dominates the size response within the first 12 hours regardless of the magnitude of drying, while the longer-term size change is gradually affected by surface drying too.

\item Given the intensity evolution, the transient response of the wind field to idealized landfalls can be predicted reasonably well by combining simple models for storm structure (Eq.\eqref{ER11theory1}-\eqref{E04theory1}) and the responses of outer size (Eq.\eqref{Eq_sizescale_theory}). These results for size and structure are expected to apply generally, as they do not depend on any free parameters.
\end{itemize}


These findings suggest that a simple theory-based model may be useful for a first-order prediction of the tropical cyclone wind field after landfall. It offers an efficient approach to generate the complete TC wind profile with limited known environmental parameters. Though systematic bias is difficult to avoid, one can use empirical adjustment to reduce or eliminate the system bias depending on the application purpose \citep{Chavas+Knaff2022}. Landfall in the real world is complicated, though. This series of studies (CC20, CC21, and the present work) removes the additional complexities existed in the real landfalls to focus on the most fundamental processes associated with landfall. The model serves as a baseline for future testing how key additional complexities, such as finite translation speed \citep{Hlywiak2021}, surface heterogeneity, and asymmetries, modify the wind field response after landfall. 

\begin{figure*}[p]
\centerline{\includegraphics[width=\textwidth]{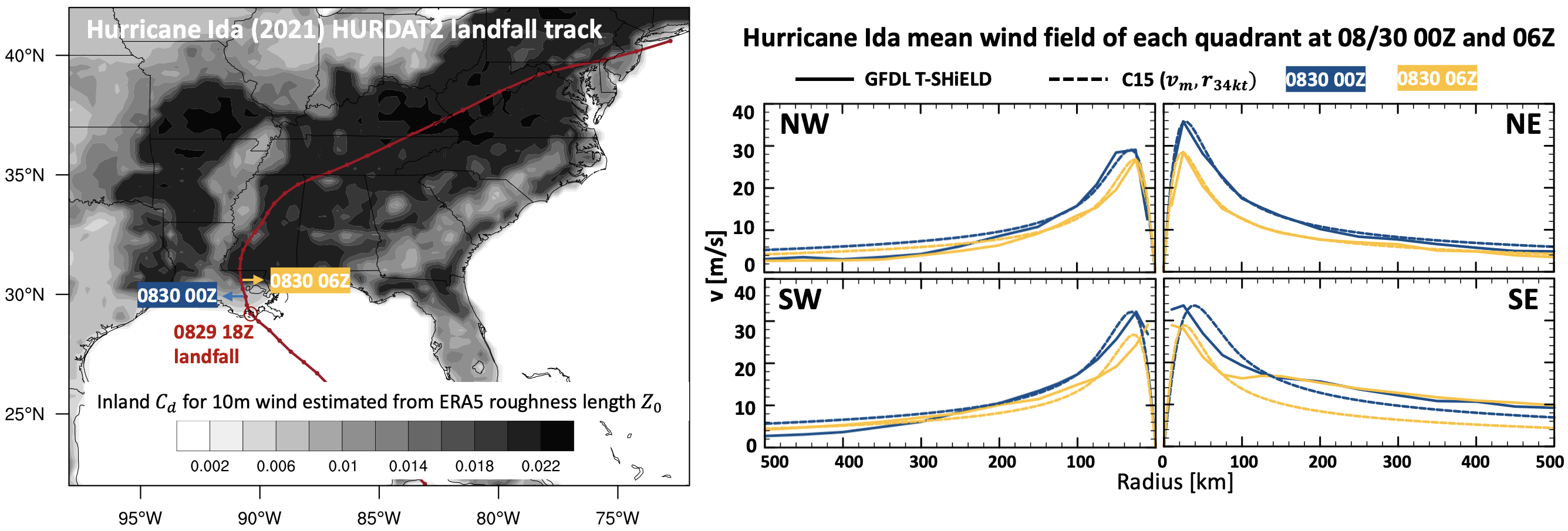}}
\caption{The azimuthally-averaged wind profile of the inland Hurricane Ida at 083000 and 083006 UTC 2021 (solid lines in the right panel) calculated from GFDL T-SHiELD model and the corresponding C15 model prediction (dash lines) using the $v_m$ and $r_{34kt}$ from T-SHiELD. In C15 prediction, $W_{cool}=0.002 ms^{-1}$ across all the quadrants; $C_d=0.02$ and $0.015$ for northeast and northwest quadrants, respectively (over the land) and $C_d=0.0015$ for southern quadrants (over the ocean) as shown on the TC track map. $C_d$ is calculated from the ERA5 surface roughness length (Eq.13 in \cite{Hersbach2010}) and then averaged within $r=500\;km$ to yield a single value within each of the four earth-relative quadrants.} \label{F8reanalysis}
\end{figure*}

Future work may test the wind field model framework against observations and reanalysis data for the real-world landfalling storms to examine the validity of the theory when applied to complicated real-world storms and environments. As a simple initial demonstration, here we show an example case-study application of our model to the Geophysical Fluid Dynamics Laboratory (GFDL) T-SHiELD real-time simulation for 2021 landfalling Hurricane Ida in Fig.\ref{F8reanalysis} \citep{Hazelton2018, Harris2020}. We take $v_m(\tau)$ and $r_{34kt}(\tau)$ from T-SHiELD; $W_{cool}$ is set as $0.002 ms^{-1}$, which is identical to idealized experiments in this work. $C_d$ is calculated from the Fifth generation of ECMWF atmospheric reanalyses of the global climate (ERA5) surface roughness \citep{Hersbach2010} and then averaged within $r=500\;km$ to yield a single value within each of the four earth-relative quadrants. Overall, our model performs reasonably well in reproducing the azimuthal wind profile in each quadrant. Future work seeks to more comprehensively examine the model against observations and simulations. Nonetheless, this example illustrates how the model could potentially be applied to generate a first-order wind field prediction of real-world storms after landfall, which is essential for improving the modeling of inland hazards both operationally and in long-term risk assessment.

\acknowledgments
The authors thank for all conversations and advice from Drs. Kerry Emanuel, Frank D. Marks, Daniel T. Dawson, and Richard H. Grant. The authors were supported by NSF grants 1826161 and 1945113. We also appreciate the feedbacks and conversations related to this research during the 35\textsuperscript{th} AMS Conference on Hurricanes and Tropical Meteorology and the Symposium on Hurricane Risk in a Changing Climate 2022. Computing resources for this work were generously supported by Purdue's Rosen Center for Advanced Computing and the Community Cluster Program \citep{McCartney2014}.

%
%
\datastatement
Datasets of relevant simulated variables from this work are archived on Purdue University Research Repository (PURR). We also provide the information needed to replicate the simulations: The model code, compilation script, and the namelist settings are available from chenjie@princeton.edu. The code for the C15 wind structure model is available at https://doi.org/doi:10.4231/CZ4P-D448.

\bibliographystyle{ametsocV6}
\bibliography{references}

\newpage


\begin{table*}[h!]
\centering
\resizebox{0.9\textwidth}{!}{

\begin{tabular}
{ |p{1.7cm} p{5cm} p{2.5cm}|p{1.7cm} p{4cm} p{2.3cm}|  }
 \hline
 \multicolumn{6}{|c|}{Basic Model Setups} \\
 \hline
 Parameter& Name& Value &Parameter& Name & Value\\
 \hline
$l_h$ &horizontal mixing length & 750 \; m    &$T_{ST}$&  surface temperature &300 \; K\\
$l_{inf}$& asymptotic vertical mixing length&  100 \; m  & $T_{tpp}$ &tropopause temperature   &200 \; K \\
$C_k$ & exchange coeff. of enthalpy &0.0015 & $Q_{cool}$& radiative cooling rate (potential temperature) & $1 \; K \: day^{-1}$\\
$C_d$ & \textbf{exchange coeff. of momentum} & \textbf{0.0015} & $idiss$& dissipative heating& 1 (turned on)\\
$\epsilon$& \textbf{surface evaporative fraction }& \textbf{1}& $f$&Coriolis Parameter& $5\times 10^{-5} s^{-1}$\\
$\Delta x$ &horizontal grid spacing&3 \; km &$H_{domain}$&height of model top& 25 \; km\\
$\Delta z$ &H=0-3\;km: fixed vertical grid spacing &0.1\;km  &$L_{domain}$&radius of model outer wall&3000 \; km\\
&H=3-12\;km: stretching vertical grid spacing &0.1 to 0.5\;km  && &\\
&H=12-25\;km: fixed vertical grid spacing &0.5\;km  && &\\
\hline 
\end{tabular}
}
\caption{Parameter values of the CM1 CTRL simulation. Only the two bold parameters $C_d$ and $\epsilon$ are modified individually or simultaneously in idealized landfall experiments as described in Fig.\ref{Table_experiments}.}
\label{geom}
\end{table*}

%

\clearpage

\clearpage

\clearpage

\clearpage

\clearpage

\clearpage

\clearpage


\clearpage

\clearpage








%



\end{document}